\documentstyle[prl,aps,multicol,epsf]{revtex}

\begin{document}
\title{An adjustable Brownian heat engine}
\author{Mesfin Asfaw and Mulugeta Bekele\cite{mulu}}
\address{Department of Physics, Addis Ababa University,
P. O. Box 1176, Addis Ababa, Ethiopia\\
and\\
The Abdus Salam International Centre for Theoretical Physics\\
11 Strada Costiera, 34014 Trieste, Italy} 
\maketitle
\begin{abstract}
   A microscopic heat engine is modeled as a Brownian particle in a sawtooh potential (with load) moving through
a highly viscous medium driven by the thermal kick it gets from alternately placed hot and cold heat reservoirs.
We found closed form expression for the current as a function of the parameters characterizing the model.
Depending on the values these model parameters take, the engine is also found to function as a refrigerator.
Expressions for the efficiency as well as for the refrigerator performance are also reported.
Study of how these quantities depend on the model parameters enabled us in identifying the points in the parameter
space where the engine performs either with maximum power or with optimized efficiency. The corresponding efficiencies of the engine
are then compared with those of the endoreversible and Carnot engines.
\end{abstract}
\centerline{PACS number(s): 05.40.Jc, 02.50.Ey, 05.40.-a, 05.60.-k, 05.70.-a}

\begin{multicols}{2}
We are in an era where devices including engines are being miniaturized to micron or even nano sizes. The physics of
such microscopic engines
is quite different from that of the macroscopic physics we daily experience. Due to their length scale, they are subjected to
overdamped Brownian motion. Any task such an engine performs, such as translocating chemicals with finite velocity, costs energy.
In order to find out how efficient such an engine is, one has to generalize the definition of efficiency. This is what Der\'enyi
{\it et al}.
did in their crucial paper on generalized efficiency \cite{derenyi}. Another issue of how an engine operates is whether a fastest
transporting velocity exists under a given condition. Such fast transport is at the expense of losing significant energy and implying
less efficiency. A compromise between fast transport and energy cost may lead to an optimized efficiency. Her\'nandez {\it et al}.
recently came up with a unified optimization method for doing exactly this \cite{calvo}.

The aim of this paper is to model an adjustable microscopic (or Brownian) heat engine that could perform tasks in different modes of
operation as mentioned above.

The idea of a Brownian heat engine working due to nonuniform temperature first came up with the works of B\"{u}ttiker, van Kampen, and
Landauer \cite{butt} while they were involved in exposing the significance of the now influential papers of Landauer on blowtorch
effect \cite{land}. Following B\"{u}ttiker's work, Matsuo and Sasa \cite{miki} considered such an engine as
an example to show that it acts as a Carnot engine at quasistatic limit. Der\'enyi and Astumian \cite{astu}, after analyzing the
details of heat flow in a Brownian heat engine, found that its efficiency can, in principle, approach that of a Carnot engine.
The same group later used a model Brownian heat engine as an example to apply their newly introduced
definition of generalized efficiency and find its implications \cite{derenyi} .

All these recent works addressed how a Brownian heat engine behaves in the quasistatic limit. Instead of taking the engine as an
example to address an issue of purely theoretical interest like quasistatic limit, we would like to take an exactly solvable model
where information of practical interest can also be extracted from the model parameters themselves. Not only will we be dealing with
the quasistatic limit but with general characteristics of real engines performing tasks either at maximum rate or at optimized efficiency
or otherwise. Moreover, some of the wrong conclusions that were given due to either overlooking or lack of exact solution will be corrected.

The model consists of a Brownian particle moving in a sawtooth potential with an external load where the viscous medium is alternately
in contact with hot and cold heat reservoirs along the space (or reaction) coordinate. The shape of a single sawtooth potential,
$U_s(x)$, located around $x=0$ is described by
\begin{equation}
  U_s(x)=\cases{
   U_{0}\left({x\over L_{1}}+1 \right),&if $-L_{1}\le x < 0$;\cr
   U_{0}\left({-x\over L_{2}}+1\right),&if $0 \le x < L_{2}$.\cr
   }
\end{equation}
\begin{figure}
\epsfxsize \columnwidth
\epsfbox{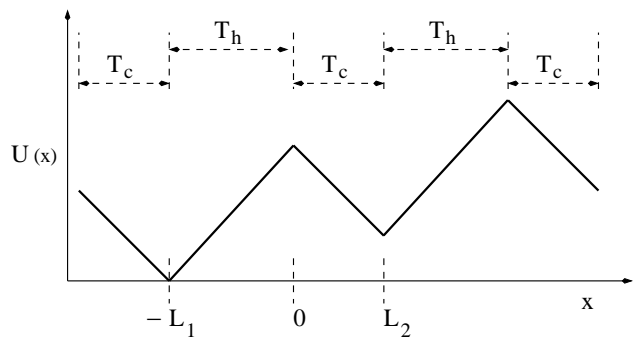}
\caption{ Plot of the sawtooth potential in the presence of constant external load. The temperature profile is shown above the potential
profile.}
\end{figure}
\noindent
The potential corresponding to the external load is linear, $fx$, where $f$ is the load.  The temperature profile,
$T(x)$, within the same interval $-L_{1}\le x < L{_2}$ is describe by
\begin{equation}
T(x)=\cases{
   T_h,&if $-L_{1}\le x < 0$;\cr
  T_c,&if $0 \le x < L_{2}$.\cr
   }
\end{equation}
Both $U_{s}(x)$ and $T(x)$ are taken to have the same period such that $U_{s}(x+L) = U_{s}(x)$ and $T(x+L)= T(x)$ where $L=L_{1}+L_{2}$.
Note that the left side of each sawtooth from its barrier top overlaps with the hot region of the medium while the right side overlaps
with the cold region. The sawtooth potential with the load, $U(x)=U_s(x) + fx$, and the temperature profile, $T(x)$, are shown in Fig. 1.

From practical point of view, the size of such microscopic heat engine is limited by how small in size temperature gradients can be.
Since temperature gradients of micron size can be produced \cite{rmp97}, the size of such an engine could be as small as few microns.

Due to the presence of the hot and cold regions within each sawtooth potential and the external load, the Brownian particle will
generally be driven unidirectionally and attain a steady state current, $J$, whose magnitude and direction depends upon the
quantities characterizing the model. The dynamic equation governing the motion of the Brownian particle in such highly viscous and
inhomogeneous medium is the corresponding Smoluchowski equation first derived by van Kampen and later by Jayannavar
and Mahato \cite{kampen88}:
\begin{equation}
{\partial \over \partial t}(P(x,t))={\partial\over  \partial x}\left [{1\over \gamma (x)}\left(U'(x)P(x,t)+{\partial \over \partial x}
       (T(x)P(x,t))
   \right)\right]
\end{equation}
where $P(x,t)$ is the probability density of finding the particle at position $x$ at time $t$, $U'(x)=dU(x)/dx$, and $\gamma(x)$
is the coefficient of friction at position $x$. Boltzmann's constant, $k_B$, is taken to be unity. The constant current at steady state
is given by
\begin{equation}
J=-{1\over \gamma (x)}\left[U'(x)P_{ss}(x)+{d \over dx}(T(x)P_{ss}(x))\right],
\end{equation}
where $P_{ss}(x)$ is the steady state probability density at position $x$. Using  periodic boundary condition, $P_{ss}(x+L)=P_{ss}(x)$,
and taking coefficient of friction the same throughout the medium,
we get a closed form expression for $J$  for our potential and temperature profiles \cite{mesfin}:
\begin{equation}
J={-F\over G_{1}G_{2} + HF},
\end{equation}
where $F$, $G_1$ and $G_2$ are
\begin{eqnarray}
F&=& e^{a-b} - 1,\nonumber \\
G_1&=&\frac{L_1}{aT_h}\left(1-e^{-a}\right) + \frac{L_2}{bT_c}e^{-a}\left(e^b - 1\right), \nonumber \\
G_2&=&\frac{\gamma L_1}{a}\left(e^a - 1\right) + \frac{\gamma L_2}{b}e^a\left(1-e^{-b}\right).
\end{eqnarray}
On the other hand, $H$ can be put as a sum of three terms: $A+B+C$, where
\begin{eqnarray}
A&=&\frac{\gamma}{T_h} \left(\frac{L_1}{a}\right)^2 (a + e^{-a}-1), \nonumber \\
B&=&\frac{\gamma L_1L_2}{abT_c} (1-e^{-a})(e^b-1), \nonumber \\
C&=& \frac{\gamma}{T_c}\left(\frac{L_2}{b}\right)^2(e^b-1-b).
\end{eqnarray}
Note that $a = (U_0 + f L_1)/T_h$ and $b = (U_0 - f L_2)/T_c)$ in Eqns. (6)
and (7).
For the case when there is no load ($f=0$) and $L_2=L_1$, the current takes a
simple expression given by
\begin{equation}
J= \frac{1}{2\gamma(T_h + T_c)}\left (\frac{U_0}{L_1}\right )^2 \left ( \frac{1}{e^{\frac{U_0}{T_h}} - 1}  -
\frac{1}{e^{\frac{U_0}{T_c}} - 1} \right ).
\end{equation}
Note that this current is a resultant of currents to the right and to the left; i.e. $J=J_+ - J_-$.

One can clearly see that the model acts as a heat engine when the current is to the right. The particle undergoes cyclic motion wherein
during each cycle it is first in contact with the hot region of width $L_1$ and then with the cold region of width $L_2$. We take
account of energy flows between the two regions neglecting energy transfer via kinetic energy due to the particle's recrossing of the
boundary between the regions \cite{astu,derenyi,valve}. As the particle moves through the hot region it absorbs a total amount of heat,
$Q_h$,
which will enable it not only to climb up the potential $(U_0 + fL_1)$ but also acquire additional energy $\gamma v L_1$ ($v$ being the
particle's drift velocity and equal to $J(L_1+L_2)$) to overcome the region's viscous drag force so that
\begin {equation}
Q_{h}=U_{0}+(\gamma v + f)L_1.
\end{equation}
\noindent
On the other hand, the cold region will absorb energy as the particle moves down the potential hill while at the same time lose some
energy due to the drag force in the region. Therefore, the net heat, $Q_c$, absorbed by the cold region will be
\begin {equation}
Q_{c}=U_{0}-(\gamma v + f)L_2.
\end{equation}
The net work, $W$, done by the engine in one cycle will then be the difference between $Q_h$ and $Q_c$ so that
\begin {equation}
W= (\gamma v+f)(L_{1}+L_{2}).
\end{equation}
Note that the term $(\gamma v + f)$ appearing in the above three expressions (Eqs. (9-11)) is the sum of the drag force due to the
particle's motion through the medium and the load being lifted. One can take this term as {\it generalized load} so that the
generalized efficiency, $\eta$,  as suggested by Der\'enyi {\it et al}. \cite{derenyi} will be given by
\begin{equation}
\eta={W\over Q_{h}}={(\gamma v+f)(L_{1}+L_{2})\over U_{0}+(\gamma v + f)L_{1}}.
\end{equation}
If, on the other hand, the load is large enough along with appropriately chosen other quantities then the current will be in the
reverse direction in which case the load does work of amount $W_{L}=f(L_1+L_2)$ in one cycle forcing heat to be extracted from the cold
region of amount $Q_{c}=U_{0}-(\gamma v + f)L_2$. Under this condition the engine acts as a refrigerator. This will then lead us to a
{\it generalized} definition of coefficient of performance (COP) of the refrigerator, $P_{ref}$, which is given by
\begin{equation}
P_{ ref}={Q_{c}\over W_{L}}={U_{0}-(\gamma v + f)L_2 \over f(L_{1}+L_{2})}.
\end{equation}
The condition at which the current changes its direction is the boundary demarcating the domain of operation of the engine as a
refrigerator from that as a heat engine. In general, this condition is satisfied when
\begin{equation}
f={U_{0}(T_{h}-T_{c}) \over L_{1}T_{c}+L_{2}T_{h}}.
\end{equation}
\noindent
It is worth noting that the magnitude of the load at this point of zero current is exactly equal to what is usually called the
{\it stall force} for molecular engines \cite{howard}. When we evaluate the expressions for both $\eta$ and $P_{ref}$ as we
approach this boundary, {\it we analytically find that they are exactly equal to the values for efficiency of the Carnot engine
and for COP of the Carnot refrigerator, respectively}: $\displaystyle \lim _{{J\to 0^{+}}}{\eta }={T_{h}-T_{c}\over   T_{h}}$
and $\displaystyle \lim _{{J\to 0^-}}{P_{ref}}={T_{c}\over T_{h}-T_{c} }$.
Hence, {\it this boundary at which current is zero corresponds to the quasistatic limit be it from the heat engine side or from the
refrigerator side}. We would like to emphasize that the efficiency of the heat engine at the stall force takes the
{\it maximum} value {\it contrary} to the claim (of zero value) made by Der\'enyi, {\it et al}. \cite{derenyi}.
We would also like to point out that the efficiency goes to zero as the barrier height, $U_0$, goes to infinity {\it contrary}
to the claim (of maximum value) made by
Der\'enyi and Astumian \cite{astu}. This can be clearly seen from Eq. (12) above. We also claim that the expression for power of a
reversible heat engine given in Eq. (10) of Der\'enyi {\it et al}.
\cite{derenyi} does not make sense since {\it one cannot have a thermal engine delivering finite power while, at the same time, operating
reversibly}.
\begin{figure}
\epsfxsize \columnwidth
\epsfbox{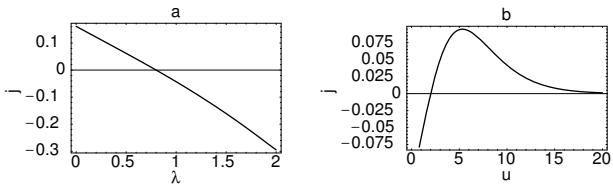}
\caption{(a) Plot of $j$ versus $\lambda$ for $\tau=1$, $\ell$=2 and $u=4$. (b) Plot of $j$ versus $u$ for $\tau=1$, $\ell$=2 and
$\lambda=0.4$.}
\end{figure}
We next explore how current, efficiency and COP of the model depend on some of the quantities characterizing it. In general,
the quantities characterizing the model are $U_0$, $L_1$, $L_2$, $f$, $T_c$ and $T_h$. We scale $U_0$, $L_2$, $T_h$ and $f$ such that
$u = U_0/T_c$, $\ell = L_2/L_1$, $\tau = (T_h/T_c) - 1$, and
$\lambda = f L_1/ T_c$.
Hence, we have four parameters $u$, $\ell$, $\tau$ and $\lambda$ characterizing the model for a given $T_c$ and $L_1$. We also scale
current such that $j=J/J_0$, where $J_0=T_c/(\gamma L_1^2)$. Figure 2(a) is a plot of the scaled current, $j$, versus
scaled load, $\lambda$. It shows that the engine works as a heat engine when the load value is less than the stall force ($\lambda=0.8$
in this case) and works as a refrigerator when the load is larger than this value. One can also see the domains of operations of
the model by plotting $j$ as a function of $u$ as has been shown in Fig. 2(b). Note that when the model works as a heat engine there is
a finite $u$ at which the current is maximum. {\it This corresponds to the point in the parameter space at which the engine operates with
maximum power}.
When we plot the efficiency, $\eta$, versus $\lambda$ within the domain where the model works as a heat engine, we find that it
increases linearly with increase in $\lambda$ attaining its maximum value (Carnot efficiency) at the stall force (See Fig. 3(a)).
On the other hand, Fig. 3(b) shows plot of $P_{ref}$ versus $\lambda$ which starts from its maximum value (COP of Carnot refrigerator) at
the stall force and decreases fast as we increase the load.
\begin{figure}
\epsfxsize \columnwidth
\epsfbox{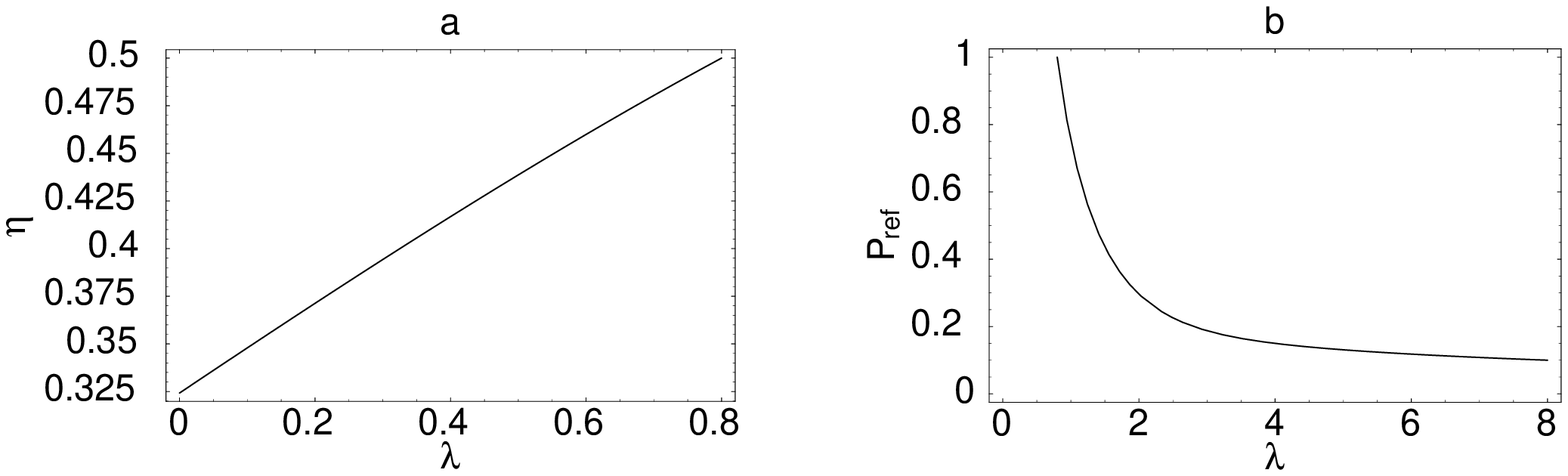}
\caption{(a) Plot of $\eta$ versus $\lambda$ for $\tau=1$, $\ell$=2 and $u= 4$. (b) Plot of $P_{ref} $ versus $\lambda$ for $\tau=1$, $\ell$=2 and $u=4$.}
\end{figure}
Let us now compare the efficiency of our model heat engine with that of the so-called endoreversible engine when both operate with
maximum power. Curzon and Ahlborn \cite{curzon} took an endoreversible engine that exchanges heat linearly at finite rate with the two
heat reservoirs and found its efficiency at maximum power, $\eta_{CA}$, to be equal to $1- \sqrt{(T_c/T_h)}$.
Figure 4 gives plots comparing $\eta_{CA}$ with that of the efficiency of our model heat engine when it also operates with maximum power,
$\eta_{MP}$, for different values of  $\tau$. The plots show that the two results are reasonably close to each other for small $\tau$
with $\eta_{CA}$ being slightly smaller than $\eta_{MP}$. The two values coincide at a particular value of $\tau$ and $\eta_{CA}$
progressively gets
larger and larger than that of $\eta_{MP}$ for higher values of $\tau$. This shows that the assumption Curzon and Ahlborn took to get
the expression for their efficiency holds reasonably well for our model heat engine as long as $\tau$ is small. It is worth
noting that our model did not take any extra assumption in the way it operates either with maximum power or otherwise. We, therefore,
believe that the reason for their deviation has got to do with the nonlinear nature of the way heat is exchanged.
\begin{figure}
\epsfxsize 7cm \epsfbox{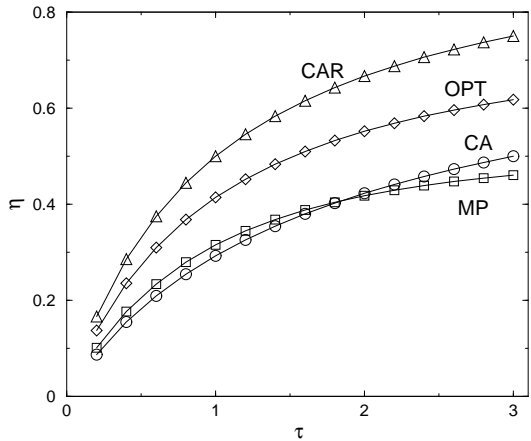}
\caption{ Plots of ${\eta}_{CA}$, ${\eta}_{MP}$, ${\eta}_{OPT}$ and ${\eta}_{CAR}$ versus $\tau$, where the model engine is put
to function at $f=0$ and $\ell=2$ while $u$ is fixed depending on whether it is working at either maximum power or optimized efficiency.}
\end{figure}
We would further like to compare Carnot efficiency, $\eta_{CAR}$, with what is called {\it optimized} efficiency, $\eta_{OPT}$, by
Her\'nandez {\it et al}. \cite{calvo}. This optimized efficiency is the efficiency at the point of operation of an engine where competition
between energy cost and fast transport is compromised. We briefly summarize the method Her\'nandez, {\it et  al}. \cite{calvo} used to get
$\eta_{OPT}$.   When the model operates with finite time the amount of work, $W$, it delivers is such that it lies
between the maximum, $W_{max}$, and minimum, $W_{min}$, amount of work that can be extracted from the model:$W_{min}\le W \le W_{max}$.
They defined two quantities: effective work, $W_e = W - W_{min}$ and noneffective work $ W_{ne}=W_{max}-W$ and introduced a function,
$\Omega$, to be optimized to be equal to the difference between these quantities. For our model heat engine (since $W_{min}=0$) this
function is given by
\begin{equation}
\Omega = 2W- \left( \frac{\tau}{1+\tau}\right )Q_h.
\end{equation}
\begin{figure}
\epsfxsize 5cm \epsfbox{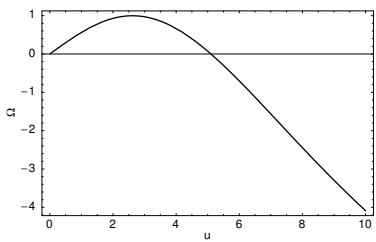}
\caption{ Plot of $\Omega $ versus $u$ for $f=0$, $\ell=2$ and $\tau=1$.}
\end{figure}
Figure 5 shows plot of $\Omega$ versus $u$ in which the function has indeed an optimum at a finite value of $u$.
The efficiency of the model heat engine when it operates at this particular point in the parameter space is what we call as optimized
efficiency, $\eta_{OPT}$. For the sake of comparison, we plot $\eta_{OPT}$ along with the corresponding Carnot efficiency, $\eta_{CAR}$,
in the same figure (Fig. 4) as we did for the other efficiencies. The plots clearly show that $\eta_{OPT}$ always lies between the
$\eta_{CAR}$ and $\eta_{MP}$. One can therefore say that {\it the operation of the model at optimized efficiency is indeed a compromise
between fast transport and energy cost}.

In conclusion, we believe that our work has, for the first time, {\it quantitatively} explored the basic properties such as current,
efficiency and COP of a model Brownian heat engine. The recently introduced definition of generalized efficiency \cite{derenyi} has been used
in getting the expressions for these quantities. We have found operating conditions of the model under which either maximum power is
delivered or optimized efficiency is attained and have been able to {\it explicitly} compare their efficiencies. These theoretical results
propose the extents and limits to be considered in the design of actual Brownian heat engines which, of course, take finite time to
accomplish any task.

We would like to thank The Abdus Salam ICTP and Swedish International
Development Agency (SIDA)
as this paper was written while the authors were visiting the Centre with their
support. We would also like to thank IPPS, Uppsala University, Sweden for the facilities they have provided for our research group.

\end{multicols}
\end{document}